\documentclass{svjour3}                     % onecolumn (standard format)
\smartqed  % flush right qed marks, e.g. at end of proof

\usepackage{url,amssymb,mathptmx,amsmath,amscd,xspace}
\DeclareFontFamily{OML}{cyi}{} \DeclareFontShape{OML}{cyi}{m}{n}{
  <5> <6> <7> <8> <9> gen * wncyi
  <10> <10.95> <12> <14.4> <17.28> <20.74> <24.88> wncyi10
 }{}
\DeclareSymbolFont{rusletters}{OML}{cyi}{m}{n}
\DeclareSymbolFontAlphabet{\rusmath}{rusletters}
\DeclareMathSymbol\re{\rusmath}{rusletters}{"03}

\providecommand{\href}[2]{#2}

\providecommand*{\eprint}[2][]{%
\href{http://arXiv.org/abs/#2}{\begingroup \Url{arXiv:#2}}%
}
\def\cprime{\/{\mathsurround=0pt$'$}}
\DeclareMathOperator{\sym}{sym}
\DeclareMathOperator{\cosym}{cosym}

\DeclareMathOperator{\id}{id}
\DeclareMathOperator{\image}{im}
\DeclareMathOperator{\kernel}{ker}
\DeclareMathOperator{\CL}{CL}
\newcommand*{\eval}[1]{\left.#1\right|}
\newcommand*{\abs}[1]{\left|#1\right|}
\newcommand{\ldb}{[\![}
\newcommand{\rdb}{]\!]}

\let\kappa=\varkappa
\let\phi=\varphi
\newcommand*{\sd}[2]{\{\,#1\mid#2\,\}}

\newcommand{\cdiff}{CDIFF\xspace}
\newcommand{\reduce}{REDUCE\xspace}
\allowdisplaybreaks[4]
\journalname{Acta Applicandae Mathematicae}
\begin{document}

\title{A unified approach to computation of integrable structures\thanks{This
    work was supported in part by the NWO-RFBR grant 047.017.015 (IK and AV),
    RFBR-Consortium E.I.N.S.T.E.IN grant 06-01-92060 (IK, AV, and RV),
    RFBR-CNRS grant 08-07-92496 (IK and AV), GNFM of Istituto Nazionale di
    Alta Matematica (IK), Monte dei Paschi di Siena through Universit\`a del
    Salento (IK and AV), Universit\`a del Salento (RV).}} %\subtitle{}

\titlerunning{A unified approach to computation of integrable structures}

\author{I.S. Krasil\cprime shchik \and A.M. Verbovetsky \and R.  Vitolo}
\institute{ I.S. Krasil\cprime shchik, A.M.  Verbovetsky, Independent
  University of Moscow, B. Vlasevsky 11, 119002 Moscow, Russia,
  \email{josephk@diffiety.ac.ru, verbovet@mccme.ru} \and R. Vitolo, Dept.\ of
  Mathematics ``E. De Giorgi'', Universit\`a del Salento, via per Arnesano,
  73100 Lecce, Italy, \email{raffaele.vitolo@unisalento.it} }

\authorrunning{I.S. Krasil\cprime shchik, A.M. Verbovetsky, R. Vitolo}

\date{Received: *** / Accepted: ***}

\maketitle

\begin{abstract}
  We expose a unified computational approach to integrable structures
  (including recursion, Hamiltonian, and symplectic operators) based on
  geometrical theory of partial differential equations.  
\end{abstract}

\section*{Introduction}
\label{sec:introduction}

Although there seems to be no commonly accepted definition of integrability for
general systems of partial differential equations
(PDEs),~\cite{Mikhailov:In,Zakharov:WIsIn}, some features of these systems are
beyond any doubt. These are:
\begin{itemize}
\item infinite hierarchies of (possibly, nonlocal) symmetries and/or
  conservation laws,
\item (bi-)Hamiltonian structures,
\item recursion operators.
\end{itemize}
Of course, these properties are closely interrelated, even if this is not so
straightforward as it may seem at first glance (see,
e.g.,~\cite{FerKhuKle:LMPh:2011}).

Methods to establish the integrability property are numerous and often reduce
to finding a Lax pair for a particular PDE. In spite of their efficiency, all
these methods possess two drawbacks that are quite serious, to our opinion:
\begin{itemize}
\item they are unable to deal with general PDEs directly and to be applied
  require reduction of the initial system to evolution form (which is not
  always done in a correct way, see, \emph{e.g.}, the discussion in
  \cite{GoKrVe-AAM-2008}),
\item they lack a well-defined language for working with nonlocal objects that
  are essential in most applications.
\end{itemize}

In a series of
papers~\cite{GoKrVe-AAM-2008,GolovkoKrasilshchikVerbovetsky:VPNSPDEq,KeKrVe-JGP-2004,KerstenKrasilshchikVerbovetsky:InCSSREvDEq,KeKrVe-AAM-2006,KerstenKrasilshchikVerbovetskyVitolo:HSGP,KerstenKrasilshchikVerbovetskyVitolo:InKD,KerstenKrasilshchikVerbovetskyVitolo:WDVV}
we presented our approach to integrable structures which is, at least
partially, free from these disadvantages. In~\cite{KrVe-JGP-2011} a review of
the geometrical theory illustrated with many examples from the above papers can
be found.

Essentially, the approach is based on two constructions naturally associated to
the equation at hand: those of tangent and cotangent coverings (see
Sec.~\ref{sec:tang-cotang-cover}) that serve as exact conceptual counterpart to
tangent and cotangent bundles in classical differential geometry. These
coverings are also differential equations, and we prove that all operators
responsible for the integrability properties (i.e., Hamiltonian, symplectic,
and recursion ones) can be identified with higher or generalized symmetries and
cosymmetries of these equations
(Sec.~\ref{sec:nonlocal-extensions}). Consequently, all computations boil down
to solving two linear equations:
\begin{equation*}
  \ell_{\mathcal{E}}(\phi)=0
\end{equation*}
and
\begin{equation*}
  \ell_{\mathcal{E}}^*(\psi)=0,
\end{equation*}
where~$\ell_{\mathcal{E}}$ is the linearization operator of the initial
equation and~$\ell_{\mathcal{E}}^*$ is its formally adjoint (see
Sec.~\ref{sec:line-ans-symm} and~\ref{sec:horiz-cohom-cons}) lifted to tangent
or cotangent coverings. The solutions that possess certain additional
properties (expressed in terms of the Schouten and Nijenhuis brackets and the
Euler operator) deliver the needed operators.

In this paper, we tried to expose our method in such a way that the interested
reader could use it as an operational manual for computations. As a tutorial
example we chose the Korteweg-de~Vries equation because all results for this
equation are well known and can be easily checked, even by hand. More
complicated equations are also briefly discussed. In particular, we will
consider equations in more than two independent variables which are naturally
presented in non-evolutionary form (Kadomtsev-Petviashvily and Plebanski
equations). The related examples appear here for the first time.  In
Sec.~\ref{sec:computer-support}, we describe a freely available software
package that was used in our computations~\cite{cdiff-man}.

\section{Differential equations and solutions}
\label{sec:diff-equat-solut}

Our general framework is based on~\cite{B-Ch-D-AMS}, but we will give a
coordinate-based exposition here.

For an equation in $n$ independent variables~$x^i$ and~$m$ unknown
functions~$u^j$ we consider the jet space~$J^\infty(n,m)$ with the
coordinates~$x^i$, $u_\sigma^j$, where~$i=1,\dots,n$, $j=1,\dots,m$
and~$\sigma=i_1i_2\dots i_{\abs{\sigma}}$, where $0\le i_k\le n$ and $i_1\le
i_2\le \cdots \le i_{\abs{\sigma}}$, is an ordered
multi-index of finite, but unlimited length~$\abs{\sigma}$. Denote
by~$\pi\colon J^\infty(n,m)\to\mathbb{R}^n$ the projection to the space of
independent variables~$x^1$, \dots, $x^n$.

The vector fields
\begin{equation}
  \label{eq:1}
  D_i=\frac{\partial}{\partial x^i}+\sum_{j,\sigma}u_{\sigma
    i}^j\frac{\partial}{\partial u_\sigma^j},\qquad i=1,\dots,n,
\end{equation}
where $\sigma i$ is the ordered multiindex obtained by $\sigma$ and $i$ after
reordering, are called \emph{total derivatives}. They define an $n$-dimensional
distribution on~$J^\infty(n,m)$ that is called the \emph{Cartan} (or
\emph{higher contact}) \emph{distribution} and is denoted
by~$\mathcal{C}$. Dually, the Cartan distribution is the annihilator of the
system of differential $1$-forms
\begin{equation}
  \label{eq:2}
  \omega_\sigma^j=\,du_\sigma^j-\sum_i u_{\sigma i}^j\,dx^i,\qquad
  j=1,\dots,m, \quad\abs{\sigma}\ge 0,
\end{equation}
the so-called \emph{Cartan}, or \emph{higher contact forms}.

Let a differential equation be given by
\begin{equation}
  \label{eq:3}
  F^l\Big(\dots,x^i,\dots,\frac{\partial^{\abs{\sigma}}u^j}{\partial
    x^\sigma},\dots\Big)=0,\qquad l=1,\dots,r.
\end{equation}
Then we consider all its differential consequences, or \emph{prolongations},
\begin{equation}
  \label{eq:4}
  D_\sigma F^l=0,\qquad l=1,\dots,r,\quad\abs{\sigma}\ge 0,
\end{equation}
where~$D_\sigma=D_{i_1}\circ\dots\circ D_{i_k}$ for~$\sigma=i_1\dots i_k$. We
take the (usually, infinite-dimensional) hypersurface~$\mathcal{E}\subset
J^\infty(n,m)$ defined by zeros of~\eqref{eq:4} as a geometrical image
of~\eqref{eq:3}.

We assume that the following two conditions hold: first, if~$G$ is a
function on~$J^\infty(n,m)$ and~$\eval{G}_{\mathcal{E}}=0$ then~$G$ is a
differential consequence of~$(F^1,\dots,F^r)$; second, for any differential
operator in total derivatives $\Delta$ such that $\Delta(F^l)=0$,
$l=1,\dots,r$, we have $\eval{\Delta}_{\mathcal{E}}=0$. The second condition
excludes gauge equations from our consideration.

Differential operators in total derivatives are said to be
$\mathcal{C}$-\emph{differential operators}.

By~\eqref{eq:4}, all total derivatives are tangent to~$\mathcal{E}$ and thus
the Cartan distribution induces an $n$-dimensional distribution
on~$\mathcal{E}$, which we call by the same name. Its $n$-dimensional integral
manifolds are solutions of~$\mathcal{E}$.

\begin{example}
  Consider the KdV equation
  \begin{equation}
    \label{eq:5}
    u_t=uu_x+u_{xxx}.
  \end{equation}
  The corresponding jet space is~$J^\infty(2,1)$ with the coordinates
  \begin{equation*}
    x,\quad t,\quad u_{i,k}=u_{\underbrace{x\dots x}_{i\
        \text{times}}\underbrace{t\dots t}_{k\ \text{times}}}
  \end{equation*}
  and the total derivatives
  \begin{equation*}
    D_x=\frac{\partial}{\partial x} + \sum_{i,k\ge
      0}u_{i+1,k}\frac{\partial}{\partial u_{i,k}},\qquad
    D_t=\frac{\partial}{\partial t} + \sum_{i,k\ge
      0}u_{i,k+1}\frac{\partial}{\partial u_{i,k}}.
  \end{equation*}
  Thus, the space~$\mathcal{E}$ is defined by the relations
  \begin{equation*}
    u_{i,k+1}=D_x^iD_t^k(u_{0,0}u_{1,0}+u_{3,0}),
  \end{equation*}
  from where it follows that all partial derivatives containing~$t$ may be
  expressed via the ones containing~$x$ only. The elements of the first group
  are called \emph{principal derivatives}, and the remaining ones are called
  \emph{parametric derivatives}~\cite{Mar:FCM:2009}.

  The functions~$x$, $t$, and the parametric derivatives~$u_i=u_{i,0}$ may be
  taken for coordinates on~$\mathcal{E}$ (internal coordinates). In terms of
  these coordinates, the total derivatives are
  \begin{equation}
    \label{eq:6}
    D_x=\frac{\partial}{\partial x} + \sum_{i\ge
      0}u_{i+1}\frac{\partial}{\partial u_i},\qquad
    D_t=\frac{\partial}{\partial t} + \sum_{i\ge
      0}D_x^i(uu_1+u_3)\frac{\partial}{\partial u_{i}},
  \end{equation}
  while the Cartan forms on~$\mathcal{E}$ are given by
  \begin{equation*}
    \omega_i=\,du_i-u_{i+1}\,dx-D_x^i(uu_1+u_3)\,dt,\qquad i\ge 0.
  \end{equation*}
\end{example}

\begin{example}\label{sec:diff-equat-solut-1}
  The KP equation
  \begin{equation}
    \label{eq:7}
    u_{yy}=u_{tx}-u_{x}^2-uu_{xx}-\frac{1}{12}u_{xxxx}
  \end{equation}
  is defined in the jet space $J^\infty(3,1)$. Its total derivatives can be
  written as
  \begin{equation*}
    D_t=\frac{\partial}{\partial t} + \sum_{\sigma\in P}
      u_{\sigma t}\frac{\partial}{\partial u_\sigma},\quad
    D_x=\frac{\partial}{\partial x} + \sum_{\sigma\in P}
      u_{\sigma x}\frac{\partial}{\partial u_\sigma},\quad
    D_y=\frac{\partial}{\partial y} + \sum_{\sigma\in P}
      u_{\sigma y}\frac{\partial}{\partial u_\sigma},
  \end{equation*}
  where $P$ is the set of multi-indexes that do not contain more than one
  instance of $y$. This is the set of multi-indexes of all parametric
  derivatives. Note that in $D_y$ the coordinates $u_{\sigma y}$, where
  $\sigma\in P$ is of the form $\sigma=\tau y$, are principal, and must be
  replaced with
  \begin{equation*}
    D_\tau\left(u_{tx}-u_{x}^2-uu_{xx}-\frac{1}{12}u_{xxxx}\right).
  \end{equation*}
\end{example}

\section{Linearization and symmetries}
\label{sec:line-ans-symm}

A (non-trivial) \emph{symmetry} of the distribution~$\mathcal{C}$
(on~$J^\infty(n,m)$ or on~$\mathcal{E}$) is a $\pi$-vertical vector field that
preserves this distribution. In other words,
\begin{equation*}
  X=\sum_{\sigma,j}a_\sigma^j\frac{\partial}{\partial u_\sigma^j}
\end{equation*}
is a symmetry if and only if
\begin{equation}
  \label{eq:8}
  [D_i,X]=0,\qquad i=1,\dots,n.
\end{equation}
Equation~\eqref{eq:8} implies that
\begin{equation*}
  a_{\sigma i}^j=D_i(a_\sigma^j),\qquad i=1,\dots,n,\quad j=1,\dots,m,\quad
  \abs{\sigma}\ge 0,
\end{equation*}
from which it follows that any symmetry is of the form
\begin{equation}
  \label{eq:9}
  \re_\phi=\sum_{\sigma,j} D_\sigma(\phi^j)\frac{\partial}{\partial u_\sigma^j}
\end{equation}
on~$J^\infty(n,m)$. Here~$\phi=(\phi^1,\dots,\phi^m)$ is an arbitrary smooth
vector function on~$J^\infty(n,m)$. Vector fields of the form~\eqref{eq:9} are
called \emph{evolutionary} and~$\phi$ is called the \emph{generating function}
(or \emph{characteristic}). Generating functions are often identified with the
corresponding symmetries and form a Lie $\mathbb{R}$-algebra with respect to
the \emph{Jacobi} (or \emph{higher contact}) \emph{bracket}
\begin{equation}
  \label{eq:10}
  \{\phi_1,\phi_2\}=\re_{\phi_1}(\phi_2)-\re_{\phi_2}(\phi_1).
\end{equation}

Given an equation~$\mathcal{E}$, its symmetry algebra~$\sym\mathcal{E}$ is
formed by the generating functions~$\phi$, satisfying the system of linear
equations (\emph{defining equations for symmetries})
\begin{equation}
  \label{eq:11}
  \ell_{\mathcal{E}}(\phi)=0,
\end{equation}
where the operator~$\ell_{\mathcal{E}}$ is defined as follows. For any smooth
vector function~$F=(F^1,\dots,F^r)$ on~$J^\infty(n,m)$ set
\begin{equation*}
  \ell_F=\left\Vert\sum_\sigma \frac{\partial F^\alpha}{\partial
    u_\sigma^\beta} D_\sigma\right\Vert.
\end{equation*}
This operator is called the \emph{linearization} of~$F$. It is a matrix
differential operator in total derivatives acting on generating functions and
producing vector functions of the same dimension as~$F$. Then, if~$\mathcal{E}$
is defined by Eq.~\eqref{eq:3} one has
\begin{equation*}
  \ell_{\mathcal{E}}=\eval{\ell_F}_{\mathcal{E}}.
\end{equation*}
The restriction of~$\ell_F$ to~$\mathcal{E}$ is well defined since~$\ell_F$ is
an operator in total derivatives.

\begin{example}
  For the KdV equation~\eqref{eq:5}, the defining equation for symmetries is
  \begin{equation}
    \label{eq:12}
    D_t(\phi)=u_1\phi+uD_x(\phi)+D_x^3(\phi),
  \end{equation}
  where~$\phi=\phi(x,t,u,u_1,\dots,u_k)$ and~$D_x$, $D_t$ are given
  by~\eqref{eq:6}.
\end{example}

\section{Conservation laws and cosymmetries}
\label{sec:horiz-cohom-cons}

Consider an equation $\mathcal{E}\subset J^\infty(n,m)$ that, for simplicity,
we will assume to be topologically trivial. This means that we assume that the
equation~\eqref{eq:3} and its prolongations~\eqref{eq:4} be topologically
trivial. This holds at least locally for most equations, and it is a reasonable
hypothesis in computations.  Define the spaces~$\Lambda_h^q(\mathcal{E})$ of
\emph{horizontal differential forms} that are generated by the elements
\begin{equation*}
  \omega=a\,dx^{i_1}\wedge\dots\wedge\,dx^{i_q},\qquad a\in
  C^\infty(\mathcal{E}).
\end{equation*}
The horizontal de~Rham differential~$d_h\colon\Lambda_h^q\to\Lambda_h^{q+1}$
is defined by
\begin{equation*}
  d_h\omega=\sum_i D_i(a)\,dx^i\wedge\,dx^{i_1}\wedge\dots\wedge\,dx^{i_q},
\end{equation*}
and~$d_h\circ d_h=0$.

An $(n-1)$-form~$\omega\in\Lambda_h^{n-1}(\mathcal{E})$ is a \emph{conservation
  law} of~$\mathcal{E}$ if~$d_h\omega=0$. A conservation law is \emph{trivial}
if it is of the form~$d_h\theta$ for
some~$\theta\in\Lambda_h^{n-2}(\mathcal{E})$. Thus we define the group of
\emph{equivalence classes of conservation laws} to be the cohomology
group~$\CL(\mathcal{E})=H_h^{n-1}(\mathcal{E})=\kernel d_h/\image d_h$.

To facilitate computation of nontrivial conservation laws, their
\emph{generating functions} (or \emph{characteristics}) are very
useful~\cite[p.\ 29]{B-Ch-D-AMS} (see also
\cite{Vin:AMS,Vin:Found,Vin:Spectr}). To present the
definition, recall that for any $\mathcal{C}$-differential
operator~$\Delta\colon V\to W$ its \emph{formal
  adjoint}~$\Delta^*\colon\hat{W}\to\hat{V}$ is uniquely defined,
where~$\hat{A}=\hom(A,\Lambda_h^n)$. Namely, for a scalar
operator~$\Delta=\sum_\sigma a_\sigma D_\sigma$ one has
\begin{equation*}
  \Delta^*=\sum_\sigma(-1)^{\abs{\sigma}}D_\sigma\circ a_\sigma
\end{equation*}
and for a matrix one~$\Delta=\Vert\Delta_{ij}\Vert$ we set
\begin{equation*}
  \Delta^*=\Vert\Delta_{ji}^*\Vert.
\end{equation*}

Now, let~$V$ be the space to which~$F=(F^1,\dots,F^r)$ belongs, $F^j$ being
the functions from equation~\eqref{eq:3}. Consider a conservation
law~$\omega\in\Lambda_h^{n-1}(\mathcal{E})$ and extend it to a horizontal
$(n-1)$-form~$\bar{\omega}$ on~$J^\infty(n,m)$. Then, due to the regularity
condition, since~$d_h(\bar{\omega})$ must vanish on~$\mathcal{E}$, we have
\begin{equation*}
  d_h(\bar{\omega})=\Delta(F)
\end{equation*}
for some $\mathcal{C}$-differential
operator~$\Delta:V\to\Lambda_h^n(J^\infty)$. The characteristic $\psi_\omega$
of~$\omega$ is defined by
\begin{equation}
  \label{eq:13}
  \psi_\omega=\eval{\Delta^*(1)}_{\mathcal{E}}.
\end{equation}
Then~$\omega$ is trivial if and only if~$\psi_\omega=0$.

\begin{example}
  Let~$\mathcal{E}$ be a system of evolution equations
  \begin{equation*}
    u_t^j=f^j(x,t,\dots u_k^\alpha,\dots),\qquad j,\alpha=1,\dots,m,
  \end{equation*}
  where subscript~$k$ indicates the number of derivatives
  over~$x$. Let~$\omega=X\,dx+T\,dt$ be a conservation law. Then its
  characteristic is
  \begin{equation*}
    \psi= \left(\frac{\delta X}{\delta u^1},\dots,\frac{\delta X}{\delta
        u^m}\right), 
  \end{equation*}
  where
  \begin{equation*}
    \frac{\delta}{\delta
      u^\alpha}=\sum_\sigma(-1)^{\abs{\sigma}}D_\sigma\circ\frac{\partial}{\partial
    u^\alpha}
  \end{equation*}
  are the \emph{variational derivatives}.
\end{example}

Generating functions satisfy the equation
\begin{equation}
  \label{eq:14}
  \ell_{\mathcal{E}}^*(\psi)=0.
\end{equation}
Arbitrary solutions of this equation are called \emph{cosymmetries}
of~$\mathcal{E}$.  The space of cosymmetries is denoted
by~$\cosym(\mathcal{E})$. Thus, construction~\eqref{eq:13} determines the
inclusion
\begin{equation}
  \label{eq:15}
  \delta\colon\CL(\mathcal{E})\to\cosym(\mathcal{E}).
\end{equation}

\begin{example}
  For the KdV equation~\eqref{eq:5} its cosymmetries are the solutions of
  \begin{equation}
    \label{eq:16}
    D_t(\psi)=uD_x(\psi)+D_x^3(\psi),
  \end{equation}
  as it follows from the expression of the adjoint operator
  of~$\ell_{\mathcal{E}}$ (see~\eqref{eq:12}).
\end{example}

\section{Nonlocal extensions}
\label{sec:nonlocal-extensions}

An invariant, geometric way to introduce nonlocal objects (e.g., variables,
operators) to the initial setting is based on the concept of \emph{differential
  coverings}~\cite[Chapter 6]{B-Ch-D-AMS}, see also the collection of
papers~\cite{KrasilshchikVinogradov:NTGDEqSCLBT}. To introduce the concept,
consider two well known examples.

\begin{example}
  Let
  \begin{equation*}
    u_t=uu_x+u_{xx}
  \end{equation*}
  be the the Burgers equation. Then the Cole-Hopf substitution
  \begin{equation*}
    u=-\frac{2v_x}{v}
  \end{equation*}
  introduces a new variable~$v$ which is nonlocally expressed in terms of the
  old one and satisfies the heat equation
  \begin{equation*}
    v_t=v_{xx}.
  \end{equation*}
\end{example}

\begin{example}
  The Miura transformation
  \begin{equation*}
    w_x=u+\frac{1}{6}w^2
  \end{equation*}
  introduces a nonlocal variable~$w$ and transforms the KdV
  equation~\eqref{eq:5} to
  \begin{equation*}
    w_t=w_{xxx}-\frac{1}{6}w^2w_x,
  \end{equation*}
  i.e., to the modified KdV equation.
\end{example}

The geometrical construction that generalizes these and similar examples is as
follows. Consider two equations~$\mathcal{E}\subset J^\infty(n,m)$
and~$\tilde{\mathcal{E}}\subset J^\infty(n,\tilde{m})$ with the same collection
of independent variables~$x^1,\dots,x^n$. A smooth
surjection~$\tau\colon\tilde{\mathcal{E}}\to\mathcal{E}$ is called a
\emph{differential covering} (or simply a \emph{covering}) of~$\mathcal{E}$
by~$\tilde{\mathcal{E}}$ if its differential takes the Cartan distribution
on~$\tilde{\mathcal{E}}$ to that on~$\mathcal{E}$. Coordinates in the fiber
of~$\tau$ are called \emph{nonlocal variables}.

In coordinates, this definition means that the total derivatives
on~$\tilde{\mathcal{E}}$ are of the form
\begin{equation*}
  \tilde{D}_i=D_i+X_i,\qquad i=1,\dots,n,
\end{equation*}
where~$X_i$ are $\tau$-vertical vector fields that satisfy the relations
\begin{equation}
  \label{eq:17}
  D_i(X_j)-D_j(X_i)+[X_i,X_j]=0,\qquad 1\le i<j\le n.
\end{equation}
Any $\mathcal{C}$-differential operator~$\Delta$ on~$\mathcal{E}$ can be lifted
to a $\mathcal{C}$-differential operator~$\tilde{\Delta}$
on~$\tilde{\mathcal{E}}$ substituting all total derivatives~$D_i$
by~$\tilde{D}_i$ .

\begin{example}\label{exmpl:cl-covs}
  Consider the case of three independent variables $x$, $y$, $z$.
  Let~$\omega=a\,dx\wedge dy+b\,dx\wedge dz$ be a conservation law
  of~$\mathcal{E}$ which has only two nontrivial components.
  Set~$\tilde{\mathcal{E}}=\mathcal{E}\times\mathbb{R}^{\infty}$ and
  \begin{align*}
    \tilde{D}_x&=D_x+w_x\frac{\partial}{\partial w}
    +w_{xx}\frac{\partial}{\partial w_x}+\cdots, \\
    \tilde{D}_y&=D_y+a\frac{\partial}{\partial w}
    +D_x(a)\frac{\partial}{\partial w_x}+\cdots, \\
    \tilde{D}_z&=D_z+b\frac{\partial}{\partial w}
    +D_x(b)\frac{\partial}{\partial w_x}+\cdots,
  \end{align*}
  where~$w$,~$w_x$,~$w_{xx},\ldots$ are coordinates in~$\mathbb{R}^{\infty}$.
  Then~$\tau\colon\mathcal{E}\times\mathbb{R}^{\infty}\to\mathcal{E}$ is a
  covering.

  The same construction works for~$n\ne3$. If~$n=2$ we get a $1$-dimensional
  covering.  For~$n>2$ it is important to remember that the initial
  conservation law must have only two nonzero components.
\end{example}

An interesting and a well known class of coverings comes from the
Wahlquist-Estabrook
construction~\cite{WahlquistEstabrook:PSNEvEq,WahlquistEstabrook:PSNEvEqII}.

\begin{example}
  For the KdV equation, consider the coverings
  \begin{equation*}
    \tilde{D}_x=D_x+X,\qquad\tilde{D}_t=D_t+T
  \end{equation*}
  such that the coefficients of the fields~$X$ and~$T$ depend on~$u_0$, $u_1$,
  $u_2$ and nonlocal variables only. Then one can show that for any such a
  covering
  \begin{gather*}
    X=u_0^2A+u_0B+C,\\
    T=\left(2u_0u_2-u_1^2+ \frac{2}{3}u_0^3\right)A+
    \left(u_2+\frac{1}{2}u_0^2\right)B+ 
    u_1[B,C]+\frac{1}{2}[B,[C,B]]+u_0[C,[C,B]]+D,
  \end{gather*}
  where the fields~$A$, $B$, $C$, and~$D$ are independent of~$u_i$ and fulfill
  the commutator relations
  \begin{gather}
    [A,B]=[A,C]=[C,D]=0,\notag\\
    [B,[B,[B,C]]]=0,\quad[B,D]+[C,[C,[C,B]]]=0,\label{eq:18}\\
    [A,D]+\frac{1}{2}[C,B]+\frac{3}{2}[B,[C,[C,B]]]=0.\notag
  \end{gather}
  Denote by~$\mathfrak{g}$ the free Lie algebra with four generators and
  defining relations~\eqref{eq:18}. Then any covering satisfying the above
  formulated ansatz is defined by a representation of~$\mathfrak{g}$ in the
  Lie algebra of vector fields on the covering.
\end{example}

The next example is crucial for the subsequent constructions.

\begin{example}\label{exam:Delta-cov}
  The class of coverings that will be introduced in this example delivers the
  solution to the following factorization problem. Assume that two
  $\mathcal{C}$-differential operators
  \begin{equation*}
    \Delta\colon V\to W,\qquad \Delta'\colon V'\to W'
  \end{equation*}
  are given, where~$V$, $W$, $V'$, $W'$ are spaces of smooth vector functions
  on~$\mathcal{E}$ of dimensions~$l$, $s$, $l'$, and~$s'$, respectively. The
  problem is: find all operators~$A\colon V\to V'$ such that there exists a
  commutative diagram
  \begin{equation*}
    \label{eq:19}
    \begin{CD}
      V@>\Delta>>W\\
      @VAVV@VVBV\\
      V'@>>\Delta'>W'
    \end{CD}
  \end{equation*}
  with some operator~$B\colon W\to W'$.
  Since the problem has obvious trivial solutions of the
  form~$A=A'\circ\Delta$, we are interested in the space
  \begin{equation}\label{eq:20}
    \mathcal{A}(\Delta,\Delta')=\frac{\sd{A}{\Delta'\circ
        A=B\circ\Delta}}{\sd{A}{A=A'\circ\Delta}}
  \end{equation}
  consisting of nontrivial solutions.  Obviously, elements of
  $\mathcal{A}(\Delta,\Delta')$ define maps from~$\ker\Delta$ to~$\ker\Delta'$.

  Assume that~$\mathcal{E}\subset J^\infty(n,m)$ with the coordinates~$x^i$,
  $u_\sigma^\alpha$. Consider the space~$J^\infty(n,m+l)$ with additional
  coordinates~$v_\sigma^\beta$. There is a natural surjection~$\xi\colon
  J^\infty(n,m+l)\to J^\infty(n,m)$, and let~$\mathcal{E}_\Delta$ be defined
  by
  \begin{equation*}
    \Delta(v)=0
  \end{equation*}
  in~$\xi^{-1}(\mathcal{E})$. Then~$\xi\colon\mathcal{E}_\Delta\to\mathcal{E}$
  is evidently a covering. We call it the $\Delta$-\emph{covering}.

  To every operator
  \begin{equation*}
    A=\left\Vert\sum_{\sigma}a_\alpha^{\alpha'\sigma}D_\sigma\right\Vert,
  \end{equation*}
  there corresponds a vector function
  \begin{equation*}
    \Phi_A=\left(\sum_{\sigma,\alpha}a_\alpha^{1,\sigma}v_\sigma^\alpha,
      \dots, \sum_{\sigma,\alpha}a_\alpha^{l',\sigma}v_\sigma^\alpha\right)
  \end{equation*}
  on~$\mathcal{E}_\Delta$.  The function~$\Phi_A$ vanishes if and only if
  $A=A'\circ\Delta$ for some operator~$A'$, so that the function~$\Phi_A$
  corresponds to the equivalence class of~$A$.

  As was indicated above, the operator~$\Delta'$ can be lifted
  to~$\tilde{\Delta}'$ in the covering under consideration.  Consider a
  function~$\Phi_A$ that satisfies the equation
  \begin{equation*}
    \tilde{\Delta}'(\Phi_A)=0.
  \end{equation*}
  It can be shown~\cite{KeKrVe-JGP-2004} that such functions are in one-to-one
  correspondence with the elements of~$\mathcal{A}(\Delta,\Delta')$ given
  by~\eqref{eq:20}.
\end{example}

\section{Tangent and cotangent coverings}
\label{sec:tang-cotang-cover}

These two coverings serve, in the geometry of differential equations, as
counterparts to tangent and cotangent bundles in differential geometry
of finite-dimensional manifolds~\cite{KrVe-JGP-2011}. Technically, both can be
obtained as particular cases of Example~\ref{exam:Delta-cov} above.

Consider an equation~$\mathcal{E}\subset J^\infty(n,m)$ and take the
operator~$\ell_{\mathcal{E}}$ for~$\Delta$ in the construction of the
$\Delta$-covering. In more detail, this means that we consider a new dependent
variable~$q=(q^1,\dots,q^m)$ and extend~$\mathcal{E}$ by the equation
\begin{equation*}
  \ell_{\mathcal{E}}(q)=0.
\end{equation*}
The resulting equation is denoted by~$\mathcal{T}(\mathcal{E})\subset
J^\infty(n,2m)$ and the
projection~$\tau\colon\mathcal{T}(\mathcal{E})\to\mathcal{E}$ is called the
\emph{tangent covering} to~$\mathcal{E}$. A important property of~$\tau$ is
that its holonomic sections, i.e.,
sections~$\phi\colon\mathcal{E}\to\mathcal{T}(\mathcal{E})$ that preserve the
Cartan distributions, are symmetries of~$\mathcal{E}$.

\begin{example}
  The tangent covering for the KdV equation is the system
  \begin{align}
    u_t&=uu_x+u_{xxx},\nonumber\\
    q_t&=u_xq+uq_x+q_{xxx}\label{eq:21}
  \end{align}
  that projects to~$\mathcal{E}$
  by~$(x,t,\dots,u_k,\dots,q_i,\dots)\mapsto(x,t,\dots,u_k,\dots)$.
\end{example}

Dually, consider the jet space~$J^\infty(n,m+r)$, where~$r$ is
from~\eqref{eq:3}, with the new dependent variable~$p=(p^1,\dots,p^r)$ and
extend~$\mathcal{E}$ by
\begin{equation*}
  \ell_{\mathcal{E}}^*(p)=0.
\end{equation*}
The equation~$\mathcal{T}^*(\mathcal{E})\subset J^\infty(n,m+r)$ together with
the natural projection~$\tau^*\colon\mathcal{T}^*(\mathcal{E})\to\mathcal{E}$
is the \emph{cotangent covering} to~$\mathcal{E}$. Its holonomic sections are
cosymmetries of~$\mathcal{E}$. Note also that~$\mathcal{T}^*(\mathcal{E})$ is
always an Euler-Lagrange equation with the Lagrangian~$L=F^1p^1+\dots+F^rp^r$.

\begin{example}\label{exmpl:KdV-cot}
  If~$\mathcal{E}$ is the KdV equation then~$\mathcal{T}^*(\mathcal{E})$ is of
  the form
  \begin{align}
    u_t&=uu_x+u_{xxx},\nonumber\\
    p_t&=up_x+p_{xxx},\label{eq:22}
  \end{align}
  cf.~\eqref{eq:16}.
\end{example}

There exist two canonical inclusions,
\begin{equation*}
  \upsilon\colon\sym(\mathcal{E})\to\CL(\mathcal{T}^*\mathcal{E})
\end{equation*}
and
\begin{equation*}
  \upsilon^*\colon\cosym(\mathcal{E})\to\CL(\mathcal{T}\mathcal{E})
\end{equation*}
(defined below), that play a very important role in practical computations
described in Sec.~\ref{sec:recurs-oper-symm}--\ref{sec:recurs-oper-cosymm}
below.

Moreover, one can prove (see \cite{KrVe-JGP-2011}) that the image
of~$\upsilon$ consists of the elements that are fiber-wise linear with respect
to the projection~$\tau\colon\mathcal{T}^*\mathcal{E}\to\mathcal{E}$. Dually,
the image of~$\upsilon^*$ coincides with conservation laws that are linear
along the fibers of the tangent covering\footnote{To be more precise, the
  images of~$\upsilon$ and~$\upsilon^*$ consist of equivalence classes that
  contain at least one fiber-wise linear element.}.

In what follows, we assume that~$r$ (the number of equations) equals~$m$ (the
number of unknown functions).

Consider a symmetry~$\phi$ of the equation~$\mathcal{E}$ and extend it to
some~$\bar{\phi}$ defined on the ambient space~$J^\infty(n,m+r)$. Then due to
the Green formula \cite[p. 191]{B-Ch-D-AMS} (see \cite[p.\ 41]{Vin:AMS} for
more details) one has
\begin{equation*}
  \langle\ell_F(\bar{\phi}),p\rangle-\langle\bar{\phi},\ell_F^*(p)\rangle =
  d_h\omega_{\bar{\phi}},
\end{equation*}
where~$\langle\cdot,\cdot\rangle$ is the natural pairing between the
spaces~$A$ and~$\hat{A}$, while the
correspondence~$\bar{\phi}\mapsto\omega_{\bar{\phi}}$ is a
$\mathcal{C}$-differential operator. We set
\begin{equation*}
  \upsilon(\phi)=\eval{\omega_{\bar{\phi}}}_{\mathcal{T}^*\mathcal{E}},
\end{equation*}
and this is a well defined operation. Note that in the right-hand side of the
above formula and in similar situations, to simplify notation, when writing a
differential form we actually mean its de Rham cohomology class.

\begin{example}
  Let~$\phi$ be a symmetry of the KdV equation. Then, due to~\eqref{eq:12}
  and~\eqref{eq:22}, one has
  \begin{multline*}
    (D_t(\phi)-D_x^3(\phi)-uD_x(\phi)-u_x\phi)p-\phi(up_x+p_{xxx}-p_t) \\
    = D_t(\phi p)-D_x\big(up\phi-3p_xD_x(\phi)+D_x^2(p\phi)\big),
  \end{multline*}
  i.e.,
  \begin{equation}\label{eq:23}
    \upsilon(\phi)=\phi p\,dx+
    \big(up\phi-3p_xD_x(\phi)+D_x^2(p\phi)\big)\,dt.
  \end{equation}
\end{example}

In a dual way, if~$\psi$ is a cosymmetry of~$\mathcal{E}$ then extend it to
some~$\bar{\psi}$ on~$J^\infty(n,2m)$, use the Green formula
\begin{equation*}
  \langle\ell_F^*(\bar{\psi}),q\rangle-\langle\bar{\psi},\ell_F(q)\rangle =
  d_h\omega_{\bar{\psi}},
\end{equation*}
and set
\begin{equation*}
  \upsilon^*(\psi)=\eval{\omega_{\bar{\psi}}}_{\mathcal{T}\mathcal{E}},
\end{equation*}
with similar considerations as $\upsilon$.

\begin{example}
  For a cosymmetry~$\psi$ of the KdV equation, using equations~\eqref{eq:16}
  and~\eqref{eq:21}, we arrive at the correspondence
  \begin{equation*}
    \upsilon^*(\psi)=\psi q\,dx+
    \big(uq\psi-3q_xD_x(\psi)+D_x^2(q\psi)\big)\,dt,
  \end{equation*}
  which in this particular case coincides with~\eqref{eq:23}.
\end{example}

\begin{example}\label{sec:nonl-forms-vect-1}
  Consider the Plebanski (or second heavenly) equation
  \begin{equation}\label{eq:24}
    u_{xz}+u_{ty}+u_{tt}u_{xx}-u_{tx}^2=0.
  \end{equation}
  Its linearization is self-adjoint (so that it is Lagrangian), hence the
  tangent covering coincides with the cotangent one; they are defined through
  the equation
  \begin{equation*}
    p_{xz}+p_{ty}+u_{tt}p_{xx}+u_{xx}p_{tt}-2u_{tx}p_{tx}=0.
  \end{equation*}

  The $u$-translation symmetry $\partial/\partial u$ has the generating
  function $\phi=1$, so we have
  \begin{multline*}
    p_{xz}+p_{ty}+u_{tt}p_{xx}+u_{xx}p_{tt}-2u_{tx}p_{tx} \\
    =D_x(p_z+u_{tt}p_x-u_{tx}p_t)+D_t(p_y+u_{xx}p_t-u_{tx}p_x)=0,
  \end{multline*}
  thus the corresponding conservation law is
  \begin{equation}
    \label{eq:25}
    \upsilon(1)= (p_y+u_{xx}p_t-u_{tx}p_x)\,dx\wedge dy\wedge dz+
            (u_{tx}p_t-p_z-u_{tt}p_x)\,dt\wedge dy\wedge dz.
  \end{equation}
\end{example}

\section{Recursion operators for symmetries}
\label{sec:recurs-oper-symm}

Consider an equation~$\mathcal{E}$ and its tangent
covering~$\tau\colon\mathcal{T}\mathcal{E}\to\mathcal{E}$.
Let~$\tilde{\ell}_{\mathcal{E}}$ be the lifting of the linearization operator
to this covering and~$\Phi=(\Phi^1,\dots,\Phi^m)$, where
\begin{equation*}
  \Phi^\alpha=\sum_{\alpha'\sigma}a_{\alpha'}^{\alpha\sigma}q_\sigma^{\alpha'},
\end{equation*}
is a solution of the equation
\begin{equation*}
  \tilde{\ell}_{\mathcal{E}}(\Phi)=0.
\end{equation*}
Then, by Example~\ref{exam:Delta-cov}, the operator
\begin{equation*}
  \mathcal{R}=
  \left\Vert\sum_{\sigma}a_{\alpha'}^{\alpha\sigma} D_\sigma\right\Vert 
\end{equation*}
takes elements of~$\ker\ell_{\mathcal{E}}$ to themselves. In other words, these
operators act on symmetries of~$\mathcal{E}$ and thus are said to be
\emph{recursion operators}.

\begin{example}
  Let
  \begin{equation*}
    u_t=u_{xx}
  \end{equation*}
  be the heat equation. Then the space of the tangent covering is given by
  \begin{equation*}
    u_t=u_{xx},\qquad q_t=q_{xx}
  \end{equation*}
  and the defining equation for recursion operators is
  \begin{equation}\label{eq:26}
    \tilde{D}_t(\Phi)=\tilde{D}_x^2(\Phi),
  \end{equation}
  where
  \begin{equation*}
    \tilde{D}_x=D_x+\sum_{i\ge 0}q_{i+1}\frac{\partial}{\partial q_i},\qquad
    \tilde{D}_t=D_t+\sum_{i\ge 0}q_{i+2}\frac{\partial}{\partial q_i}
  \end{equation*}
  (recall that $q_i=q_{x\cdots x}$, $i$ times $x$) and
  \begin{equation*}
    \Phi=a^0q+a^1q_1+\dots+a^kq_k,\qquad a_i\in C^\infty(\mathcal{E}).
  \end{equation*}
  The first three nontrivial solutions of~\eqref{eq:26} are
  \begin{equation*}
    \Phi_{00}=q,\qquad\Phi_{10}=q_1,\qquad\Phi_{11}=tq_1+\frac{x}{2},
  \end{equation*}
  to which there correspond the recursion operators
  \begin{equation*}
    \mathcal{R}_{00}=\id \qquad \mathcal{R}_{10}=D_x, \qquad
    \mathcal{R}_{11}=tD_x+\frac{x}{2}.
  \end{equation*}
  It can be shown that the entire algebra of recursion operators for the heat
  equation is generated by~$\mathcal{R}_{10}$ and~$\mathcal{R}_{11}$ with the
  only relation
  \begin{equation*}
    [\mathcal{R}_{10},\mathcal{R}_{11}]=\frac{1}{2}.
  \end{equation*}
  This algebra is isomorphic to the universal enveloping algebra for the
  $3$-dimensional Heisenberg algebra.
\end{example}

\begin{example}
  Taking the tangent covering~\eqref{eq:21} to the KdV equation and
  solving~$\tilde{\ell}_{\mathcal{E}}(\Phi)=0$ in this covering leads to the
  only solution~$\Phi=q$, i.e., it yields the identity operator. This is not a
  surprise, because it is well known that the KdV equation does not admit
  local recursion operators.

  Let us now take a cosymmetry~$\psi=1$ of the KdV equation and consider the
  corresponding conservation law on the tangent covering
  \begin{equation*}
    \upsilon^*(1)=q\,dx+(uq+q_2)\,dt.
  \end{equation*}
  Then the corresponding nonlocal variable  (see
  Example~\ref{exmpl:cl-covs}) $Q^1$ satisfies the equations
  \begin{equation*}
    Q_x^1=q,\qquad Q_t^1=qu+q_{xx}.
  \end{equation*}
  In the extended setting the linearized
  equation~$\tilde{\ell}_{\mathcal{E}}(\Phi)=0$, where
  \begin{equation*}
    \Phi=A_1Q^1+a^0q+a^1q_1+\dots+a^kq_k,
  \end{equation*}
  acquires a new nontrivial solution
  \begin{equation*}
    \Phi_1=q_2+\frac{2}{3}uq+\frac{1}{3}u_1Q^1
  \end{equation*}
  to which the Lenard recursion operator
  \begin{equation}\label{eq:27}
    \mathcal{R}=D_x^2+\frac{2}{3}u+\frac{1}{3}u_1D_x^{-1}
  \end{equation}
  corresponds.

  If we consider the next cosymmetry~$\psi=u$ then the corresponding nonlocal
  variable will be defined by
  \begin{equation*}
    Q_x^3=qu,\qquad Q_t^3=qu^2+q_{xx}u-q_xu_x+qu_{xx},
  \end{equation*}
  which will lead to another recursion operator
  \begin{equation*}
    D_x^4+\frac{4}{3}uD_x^2+2u_1D_x+ \frac{4}{9}(u^2+3u_2)
  +\frac{1}{3}(uu_1+u_3)D_x^{-1}+\frac{1}{9}u_1D_x^{-1}\circ u,
  \end{equation*}
 the square of~\eqref{eq:27}.
\end{example}

Given an equation~$\mathcal{E}$ and two recursion operators~$\mathcal{R}_1$,
$\mathcal{R}_2\colon\sym\mathcal{E}\to\sym\mathcal{E}$, the Nijenhuis bracket
\begin{equation*}
  \ldb\mathcal{R}_1,\mathcal{R}_2\rdb\colon
  \sym\mathcal{E}\times\sym\mathcal{E}\to\sym\mathcal{E}
\end{equation*}
is defined by
\begin{multline*}
  \ldb\mathcal{R}_1,\mathcal{R}_2\rdb(\phi_1,\phi_2) =
  \{\mathcal{R}_1(\phi_1),\mathcal{R}_2(\phi_2)\} +
  \{\mathcal{R}_2(\phi_1),\mathcal{R}_1(\phi_2)\} \\ -
  \mathcal{R}_1(\{\mathcal{R}_2(\phi_1),\phi_2\} +
  \{\phi_1,\mathcal{R}_2(\phi_2)\}) -
  \mathcal{R}_2(\{\mathcal{R}_1(\phi_1),\phi_2\} +
  \{\phi_1,\mathcal{R}_1(\phi_2)\}) \\ + (\mathcal{R}_1\circ\mathcal{R}_2 +
  \mathcal{R}_2\circ\mathcal{R}_1)\{\phi_1,\phi_2\},
\end{multline*}
where~$\{\cdot,\cdot\}$ is the Jacobi bracket given by
Equation~\eqref{eq:10}. An operator~$\mathcal{R}$ is called \emph{hereditary}
if~$\ldb\mathcal{R},\mathcal{R}\rdb=0$.  This property was introduced
in~\cite{FoFu80,Fuch79,FuFo81,Mag80}.  It can be shown that if this property
holds and~$\mathcal{R}$ is invariant with respect to a symmetry~$\phi$ then the
symmetries~$\phi_i=\mathcal{R}^i(\phi)$ form a commuting hierarchy (see, for
example,~\cite[Ch.\ 4, Sec.\ 3.1]{KeKr00}).

\section{Symplectic structures}
\label{sec:sympl-struct}

Let us now solve the equation
\begin{equation*}
  \tilde{\ell}_{\mathcal{E}}^*(\Psi)=0
\end{equation*}
on the tangent covering for~$\Psi$ of the form
\begin{equation*}
  \Psi=(\Psi^1,\dots,\Psi^r),\qquad
  \Psi^\alpha=\sum_{\alpha'\sigma}a_{\alpha'}^{\alpha\sigma}q_\sigma^{\alpha'}.
\end{equation*}
The corresponding $\mathcal{C}$-differential operator
\begin{equation*}
  \mathcal{S}=\left\Vert\sum_{\sigma}a_{\alpha'}^{\alpha\sigma}D_\sigma\right\Vert,
\end{equation*}
due to the results of Example~\ref{exam:Delta-cov}, takes symmetries
of~$\mathcal{E}$ to its cosymmetries.

Let us call a conservation law~$\omega$ \emph{admissible} (with respect
to~$\mathcal{S}$) if
\begin{equation*}
  \delta(\omega)\in\image\mathcal{S},
\end{equation*}
where the operator~$\delta\colon\CL(\mathcal{E})\to\cosym(\mathcal{E})$ is
from~\eqref{eq:15}. Then~$\mathcal{S}$ determines a bracket on the set of
admissible conservation laws by
\begin{equation}\label{eq:28}
  \{\omega,\omega'\}_{\mathcal{S}}=L_{\re_\phi}(c'),
\end{equation}
where $L_{\re_\phi}(c')=\ell_{c'}(\phi)$, $\phi$ is such
that~$\delta(\omega)=\mathcal{S}(\phi)$ and $c'$ is a representative of the
cohomology class $\omega'$. This bracket is well defined on
equivalence classes of nontrivial admissible conservation laws. It is
skew-symmetric if
\begin{equation}
  \label{eq:29}
  \mathcal{S}^*\circ\ell_{\mathcal{E}}=\ell_{\mathcal{E}}^*\circ\mathcal{S},
\end{equation}
i.e., if~$\ell_{\mathcal{E}}^*\circ\mathcal{S}$ is a self-adjoint operator. In
the case of evolution equations,~\eqref{eq:29} means the~$\mathcal{S}$ is a
skew-adjoint operator.

Let now an operator~$\mathcal{S}$ satisfy Eq.~\eqref{eq:29}. Let us
deduce the conditions under which the bracket~\eqref{eq:28} fulfills the Jacobi
identity. To this end, consider a vector function~$\phi=(\phi^1,\dots,\phi^m)$
on~$J^\infty(n,m)$. Then, since~$\mathcal{E}$ satisfies the regularity
condition from Sec.~\ref{sec:diff-equat-solut}, Equality~\eqref{eq:29} implies
\begin{equation*}
  \ell_F^*\mathcal{S}(\phi)-\mathcal{S}^*\ell_F(\phi)=\bar{\Delta}_\phi(F),
\end{equation*}
where~$\bar{\Delta}_{\phi}$ is a $\mathcal{C}$-differential
operator. Put~$\Delta_\phi=\eval{\bar{\Delta}_{\phi}}_{\mathcal{E}}$ and
define the map
\begin{equation*}
  \delta\mathcal{S}\colon\sym\mathcal{E}\times\sym\mathcal{E}\to
  \cosym\mathcal{E}
\end{equation*}
by
\begin{equation*}
  (\delta\mathcal{S})(\phi_1,\phi_2)=(\re_{\phi_1}\mathcal{S})(\phi_2) -
  (\re_{\phi_2}\mathcal{S})(\phi_1) + \Delta_{\phi_2}^*(\phi_1).
\end{equation*}
Then the equality
\begin{equation}
  \label{eq:30}
  \delta\mathcal{S}=0
\end{equation}
guarantees the Jacobi identity for the
bracket~$\{\cdot,\cdot\}_{\mathcal{S}}$. Operators satisfying~\eqref{eq:29}
and~\eqref{eq:30} are called \emph{symplectic structures}.

\begin{example}
  The KdV equation does not possess local symplectic structures, but if we
  extend the space~$\mathcal{T}\mathcal{E}$ with nonlocal variables~$Q^1$
  and~$Q^3$ (see Section~\ref{sec:recurs-oper-symm}) and solve the equation
  \begin{equation*}
    \tilde{\ell}_{\mathcal{E}}(\Psi)=0
  \end{equation*}
  for~$\Psi$ of the form
  \begin{equation*}
    \Psi=A_3Q^3+A_1Q^1+a^0q+a^1q_1+\dots+a^kq_k,
  \end{equation*}
  there will be two nontrivial solutions
  \begin{equation*}    
  \Psi_1= Q_1,\qquad
  \Psi_3= q_1+\frac{1}{3}uQ_1+\frac{1}{3}Q_3
  \end{equation*}
  with the corresponding nonlocal symplectic operators
  \begin{equation*}
    \mathcal{S}_1=D_x^{-1}
  \end{equation*}
  and
  \begin{equation*}
    \mathcal{S}_3=D_x+\frac{1}{3}uD_x^{-1}+\frac{1}{3}D_x^{-1}\circ u.
  \end{equation*}
\end{example}

\section{Hamiltonian structures}
\label{sec:hamilt-struct}

Consider now the cotangent covering to~$\mathcal{E}$ and the
lifting~$\tilde{\ell}_{\mathcal{E}}$ of the linearization operator to this
covering. If~$\Phi=(\Phi^1,\dots,\Phi^m)$ is of the form
\begin{equation*}
  \Phi^\alpha=\sum_{\alpha'\sigma}a_{\alpha'}^{\alpha\sigma}q_\sigma^{\alpha'}
\end{equation*}
and satisfies
\begin{equation*}
  \tilde{\ell}_{\mathcal{E}}(\Phi)=0
\end{equation*}
then, due to Example~\ref{exam:Delta-cov}, the operator
\begin{equation*}
  \mathcal{H}=\left\Vert\sum_{\sigma}a_{\alpha'}^{\alpha\sigma}D_\sigma\right\Vert
\end{equation*}
takes cosymmetries of the equation~$\mathcal{E}$ to its symmetries.

For any such an operator, consider the bracket
\begin{equation}
  \label{eq:31}
  \{\omega,\omega'\}_{\mathcal{H}}=L_{\mathcal{H}(\delta\omega)}(\omega')
\end{equation}
defined on the space~$\CL(\mathcal{E})$ of conservation
laws. Here~$\delta\colon\CL(\mathcal{E})\to\cosym(\mathcal{E})$ is the
map~\eqref{eq:15}. This bracket is skew-symmetric if
\begin{equation}
  \label{eq:32}
  (\ell_{\mathcal{E}}\circ\mathcal{H})^*=\ell_{\mathcal{E}}\circ\mathcal{H}.
\end{equation}

  For evolution equations one usually requires an additional condition of
  existence of a conservation law~$\chi$ such that
  $u_t=\mathcal{H}(\delta\chi)$. For such equations~\eqref{eq:32} implies
  $\mathcal{H}^*=-\mathcal{H}$.

For any two $\mathcal{C}$-differential operators~$\mathcal{H}_1$,
$\mathcal{H}_2\colon\cosym(\mathcal{E})\to\sym(\mathcal{E})$
satisfying~\eqref{eq:32} define their Schouten
bracket~\cite[p.\ 13]{KrVe-JGP-2011}
(see~\cite[p.\ 226]{Vin:AMS} for a more general definition)
\begin{equation*}
  \ldb\mathcal{H}_1,\mathcal{H}_2\rdb\colon
  \cosym(\mathcal{E})\times\cosym(\mathcal{E})\to\sym(\mathcal{E})
\end{equation*}
by
\begin{multline}
  \label{eq:33}
  \ldb\mathcal{H}_1,\mathcal{H}_2\rdb(\psi_1,\psi_2) =
  \mathcal{H}_1(L_{\mathcal{H}_2\psi_1}(\psi_2)) -
  \mathcal{H}_2(L_{\mathcal{H}_1\psi_1}(\psi_2)) \\+
  \{\mathcal{H}_1(\psi_2),\mathcal{H}_2(\psi_1)\} -
  \{\mathcal{H}_1(\psi_1),\mathcal{H}_2(\psi_2)\},
\end{multline}
where~$\{\cdot,\cdot\}$ is the Jacobi bracket~\cite[p.\ 48]{B-Ch-D-AMS} and
\begin{equation*}
  L_\phi(\psi)=\re_\phi(\psi)+\ell_\phi^*(\psi)
\end{equation*}
for any~$\phi\in\sym(\mathcal{E})$ and~$\psi\in\cosym(\mathcal{E})$.

An operator~$\mathcal{H}$ satisfying~\eqref{eq:32} and such that
\begin{equation*}
  \ldb\mathcal{H},\mathcal{H}\rdb=0
\end{equation*}
is called a \emph{Hamiltonian structure} on~$\mathcal{E}$. For such a
structure, the bracket~\eqref{eq:31} satisfies the Jacobi identity. Two
Hamiltonian structures $\mathcal{H}_1$, $\mathcal{H}_2$, are compatible if
\begin{equation*}
  \ldb\mathcal{H}_1,\mathcal{H}_2\rdb=0.
\end{equation*}
Equations admitting compatible structures are called bi-Hamiltonian. One can
apply the Magri scheme~\cite{Magri:SMInHEq} to generate infinite series of
commuting symmetries and conservation laws.

\begin{example}
  In the case of the KdV equation, we solve the equation
  \begin{equation*}
    \tilde{D}(\Phi)=u_1\Phi+u\tilde{D}_x(\Phi)+\tilde{D}_x^3(\Phi),\qquad
    \Phi=a^0p+a^1p_1+\dots+a^kp_k,
  \end{equation*}
  in the cotangent covering (Example~\ref{exmpl:KdV-cot}) and obtain two
  nontrivial solutions
  \begin{equation*}
    \Phi_1 = p_1,\qquad
    \Phi_3 = p_3+\frac{2}{3}up_1+\frac{1}{3}u_1p_0
  \end{equation*}
  with the corresponding (and well known!) Hamiltonian operators
  \begin{equation*}
    \mathcal{H}_1 = D_x,\qquad
    \mathcal{H}_3 = D_x^3+\frac{2}{3}uD_x+\frac{1}{3}u_1.
  \end{equation*}

  Consider the $x$-translation~$u_x$ which is a symmetry of the KdV equation
  and the corresponding nonlocal variable~$P^1$ in the cotangent covering. We
  have
  \begin{equation*}
    P_x^1=pu_x,\qquad
    P_t^1=p(uu_1+u_3)+p_2u_1-p_1u_2.
  \end{equation*}
  One obtains a new solution
  \begin{equation*}
    \Phi_5 = p_5+\frac{4}{3}up_3+ 2u_1p_2+\frac{4}{9}(u^2+3u_2)p_1
    +(\frac{4}{9}uu_1+\frac{1}{3}u_3)p-\frac{1}{9}u_1P_1
  \end{equation*}
  in the extended setting to which the nonlocal Hamiltonian structure
  \begin{equation*}
    \mathcal{H}_5= D_x^5+\frac{4}{3}uD_x^3+ 2u_1D_x^2+
    \frac{4}{9}(u^2+3u_2)D_x
    +(\frac{4}{9}uu_1+\frac{1}{3}u_3)-\frac{1}{9}u_1D_x^{-1}\circ u_1
  \end{equation*}
  corresponds.
\end{example}

\begin{example}
  Consider the KP equation (Example~\ref{sec:diff-equat-solut-1}). The
  cotangent covering is defined through the further equation
  \begin{equation*}
    p_{yy}=p_{tx}-up_{xx}-\frac{1}{12}p_{xxxx}.
  \end{equation*}
  We will solve the equation
  \begin{equation*}
    \tilde{\ell}_{\mathcal{E}}(\Phi)=\tilde{D}_{yy}(\Phi)
    -\tilde{D}_{tx}(\Phi)+2u_{x}\tilde{D}_x(\Phi)
    +u_{xx}\Phi+u \tilde{D}_{xx}(\Phi)+\frac{1}{12}\tilde{D}_{xxxx}(\Phi)=0,
  \end{equation*}
  with $\Phi=\sum_{\sigma}a^\sigma p_\sigma$ (here the sum extends to a finite
  number of multi-indexes of parametric coordinates $\sigma\in P$, and
  $a^\sigma$ are smooth functions on~$\mathcal{E}$).  The nontrivial solution
  \begin{equation*}
    \Phi=p_{xx}
  \end{equation*}
  can be found. A simple computation shows that the corresponding
  operator~$\mathcal{H}=D_x^2$ satisfies~\eqref{eq:32} and fulfills the
  condition $\ldb\mathcal{H},\mathcal{H}\rdb=0$, hence it is Hamiltonian.

  In~\cite{Kupershmidt:KP} the following evolution form of the KP equation is
  introduced:
  \begin{equation}\label{eq:34}
    \psi_x=u_y,\quad u_x=v,\quad v_x=w,\quad w_x=12(u_t-uv-\psi_y).
  \end{equation}
  Then, one local Hamiltonian operator (Eq.~(51) in~\cite{Kupershmidt:KP}) is
  obtained. After applying the change of coordinate formulae for Hamiltonian
  operators (see~\cite{KerstenKrasilshchikVerbovetskyVitolo:HSGP}) it is
  readily proved that it corresponds to the above $\Phi$.
\end{example}

\section{Recursion operators for cosymmetries}
\label{sec:recurs-oper-cosymm}

Finally, let us consider the equation~$\ell_{\mathcal{E}}^*(\Psi)=0$ lifted to
the cotangent covering. An operator $\Psi$ that fulfills the equation takes
an element~$\psi\in\cosym(\mathcal{E})$ into an element
$\Psi(\psi)\in\cosym(\mathcal{E})$, i.e., it is a \emph{recursion
  operators for cosymmetries}.

\begin{example}
  If~$\mathcal{E}$ is the KdV equation then the equation
  \begin{equation*}
    \tilde{D}_t(\Psi)=u\tilde{D}_x(\Psi)+\tilde{D}_x^3(\psi)
  \end{equation*}
  possesses no nontrivial solution. But when we
  extend~$\mathcal{T}^*\mathcal{E}$ by the nonlocal variable~$P^1$ (see the
  previous section) a solution
  \begin{equation*}
    \Psi=p_2+\frac{2}{3}up-\frac{1}{3}P^1
  \end{equation*}
  arises to which the recursion operator
  \begin{equation*}
    \bar{\mathcal{R}}=D_x^2+\frac{2}{3}u-D_x^{-1}\circ u_1
  \end{equation*}
  corresponds.
\end{example}

\begin{example}
  Consider the Plebanski equation (Example~\ref{sec:nonl-forms-vect-1}).  The
  linearization of this equation is self-adjoint, hence the operators that we
  defined in this paper (recursion for symmetries, symplectic, Hamiltonian,
  recursion for cosymmetries) act in the same spaces. Now, extend the
  (co)tangent covering with the nonlocal variable~$R$ defined by
  \begin{equation*}
    R_t=u_{tx}p_t-p_z-u_{tt}p_x,\quad R_x=p_y+u_{xx}p_t-u_{tx}p_x.
  \end{equation*}
  The extension of the equation~$\tilde{\ell}_{\mathcal{E}}(\Phi)=0$ admits the
  solution linear with respect to $p_\sigma$ and $R_\sigma$
  \begin{equation}
    \label{eq:35}
    \Phi_2 = R,
  \end{equation}
  besides the trivial solution $\Phi_1=p$.

  In~\cite{NeyziNutkuSheftel:JPA:2005} the following evolution form of
  Plebanski equation is introduced:
  \begin{equation}\label{eq:36}
    u_t=q,\qquad q_t=\frac{1}{u_{xx}}(q_x^2-q_y-u_{xz}).
  \end{equation}
  Then, a local Hamiltonian operator (Eq.~(11)
  in~\cite{NeyziNutkuSheftel:JPA:2005}) and a nonlocal Hamiltonian operator
  (Eq.~(27) in~\cite{NeyziNutkuSheftel:JPA:2005}) are obtained. After applying
  the change of coordinate formulae for Hamiltonian operators
  (see~\cite{KerstenKrasilshchikVerbovetskyVitolo:HSGP}) it is readily proved
  that they correspond to $\Phi_1$ and $\Phi_2$. Different changes of
  coordinates from~\eqref{eq:24} to~\eqref{eq:36} transform~$\Phi_1$
  and~$\Phi_2$ to recursion operators for symmetries or cosymmetries and
  symplectic operators.

  It is interesting to remark that, while in the evolutionary form all such
  operators are mutually different because they act in different spaces, in the
  original formulation of the equation they coincide with $\Phi_1$ and
  $\Phi_2$, whose expressions are considerably simpler than that of their
  `evolutionary' counterparts.
\end{example}

\section{Computer support}
\label{sec:computer-support}

The computations in this paper were done by the help of \cdiff, a symbolic
computation package devoted to computations in the geometry of DEs and
developed by P.K.H.~Gragert, P.H.M.~Kersten, G.F.~Post and G.H.M.~Roelofs at
the University of Twente, The Netherlands, with latest additions by one of us
(R.V.). \cdiff runs in the computer algebra system \reduce, which recently has
become free software and can be downloaded here~\cite{red}.

The `Twente' part of \cdiff package is included in the official \reduce
distribution, and most of it is documented. Additional software, especially
geared towards computations for differential equations with more than two
independent variables, can be downloaded at the Geometry of Differential
Equations website \url{http://gdeq.org}.  The software includes a user guide
\cite{cdiff-man} and many example programs which cover most examples presented
in this paper.

\textbf{Acknowledgments.} We thank the referees for useful comments that helped
us to improve the text.


\begin{thebibliography}{99}

\bibitem{B-Ch-D-AMS} A.V.~Bocharov, V.N.~Chetverikov, S.V.~Duzhin, et al.,
  Symmetries and Conservation Laws for Differential Equations of Mathematical
  Physics, xiv+333~pp. Amer.\ Math.\ Soc., Providence, RI (1999). Edited and
  with a preface by I.~Krasil{\cprime}shchik and A.~Vinogradov.

\bibitem{FerKhuKle:LMPh:2011} E.V.~Ferapontov, K.R.~Khusnutdinova, and
  C.~Klein, \emph{On linear degeneracy of integrable quasilinear systems in
    higher dimensions}, Lett.\ Math.\ Phys.\ \textbf{96} (2011), 5--35,
  \eprint{0909.5685}.

\bibitem{FoFu80} A. S. Fokas and B. Fuchssteiner, \emph{On the structure of
    symplectic operators and hereditary symmetries}, Lett. Nuovo Cimento (2),
  \textbf{28} no. 8 (1980),
  299--303. \url{http://fuchssteiner.info/papers/30.pdf}

\bibitem{Fuch79} B. Fuchssteiner, \emph{Application of hereditary symmetries to
    nonlinear evolution equations}, Nonlinear Analysis, Theory, Methods \&
  Applications, \textbf{3} no.\ 11 (1979), 849--862.
  \url{http://fuchssteiner.info/papers/25.pdf}

\bibitem{FuFo81} B. Fuchssteiner and A. S. Fokas, \emph{Symplectic structures,
    their B\"acklund transformations and hereditary symmetries}, Phys. D
  \textbf{4} no.\ 1 (1981), 47--66
  \url{http://fuchssteiner.info/papers/38.pdf}.

\bibitem{Mag80} F. Magri, \emph{A geometrical approach to the nonlinear
    solvable equations}, Lecture Notes in Physics \textbf{120} (1980),
  Springer-Verlag, 233--263.

\bibitem{GoKrVe-AAM-2008} V.A.~Golovko, P.H.M. Kersten,
  I.S.~Krasil{\cprime}shchik, and A.M.~Verbovetsky, \emph{On integrability of
    the Camassa-Holm equation and its invariants}, Acta Appl.\ Math.\
  \textbf{101} (2008), 59--83, \url{arXiv:0812.4681}.

\bibitem{GolovkoKrasilshchikVerbovetsky:VPNSPDEq} V.A.~Golovko,
  I.S.~Krasil{\cprime}shchik, and A.M. Verbovetsky, \emph{Variational
    {P}oisson-{N}ijenhuis structures for partial differential equations},
  Theoret.\ and Math.\ Phys. \textbf{154} (2008), 227--239,
  \eprint{0812.4684}.

\bibitem{KeKr00} I.S. Krasil{\cprime}shchik, P.H.M. Kersten, Symmetries and
  recursion operators for classical and supersymmetric differential equations,
  Kluwer, 2000.

\bibitem{KeKrVe-JGP-2004} P.~Kersten, I.~Krasil{\cprime}shchik, and
  A.~Verbovetsky, \emph{Hamiltonian operators and $\ell^*$-coverings}, J.\
  Geom.\ Phys.\ \textbf{50} (2004), 273--302, \url{arXiv:math/0304245}.

\bibitem{KerstenKrasilshchikVerbovetsky:InCSSREvDEq} P.~Kersten,
  I.~Krasil{\cprime}shchik, and A.~Verbovetsky, \emph{On the integrability
    conditions for some structures related to evolution differential
    equations}, Acta Appl.\ Math.\ \textbf{83} (2004), 167--173,
  \eprint{math/0310451}.

\bibitem{KeKrVe-AAM-2006} P.~Kersten, I.~Krasil{\cprime}shchik, and
  A.~Verbovetsky, \emph{A geometric study of the dispersionless Boussinesq
    type equation}, Acta Appl.\ Math.\ \textbf{90} (2006), 143--178,
  \url{arXiv:nlin/0511012}.

\bibitem{KerstenKrasilshchikVerbovetskyVitolo:HSGP} P.~Kersten,
  I.~Krasil{\cprime}shchik, A.~Verbovetsky, and R.~Vitolo, \emph{Hamiltonian
    structures for general {PDE}s}, Differential equations: Geometry,
  Symmetries and Integrability. The Abel Symposium 2008 (B.~Kruglikov,
  V.~V. Lychagin, and E.~Straume, eds.), Springer-Verlag, 2009, pp.~187--198,
  \eprint{0812.4895}.

\bibitem{KerstenKrasilshchikVerbovetskyVitolo:InKD} P.H.M.~Kersten,
  I.S.~Krasil{\cprime}shchik, A.M.~Verbovetsky, and R.~Vitolo,
  \emph{Integrability of {K}upershmidt deformations}, Acta Appl.\
  Math. \textbf{109} (2010), 75--86, \eprint{0812.4902}.

\bibitem{KerstenKrasilshchikVerbovetskyVitolo:WDVV} P.H.M.~Kersten,
  I.S.~Krasil{\cprime}shchik, A.M.~Verbovetsky, and R.~Vitolo, \emph{On
  integrable structures for a generalized Monge-Amp\`ere equation},
  \eprint{1104.0258}.

\bibitem{KrVe-JGP-2011} I.~Krasil{\cprime}shchik and A.~Verbovetsky,
  \emph{Geometry of jet spaces and integrable systems}, J.\ Geom.\ Phys.\
  \textbf{61} (2011), 1633--1674, \url{arXiv:1002.0077}.
  
\bibitem{KrasilshchikVinogradov:NTGDEqSCLBT} I.S.~Krasil{\cprime}shchik and
  A.M.~Vinogradov, \emph{Nonlocal trends in the geometry of differential
    equations: {S}ymmetries, conservation laws, and {B}{\"a}cklund
    transformations}, Acta Appl.\ Math. \textbf{15} (1989), 161--209.

\bibitem{Kupershmidt:KP} B.A.~Kupershmidt, \emph{Geometric Hamiltonian forms
    for the Kadomtsev-Petviashvily and Zabolotskaya-Khokhlov equations}, in
  Geometry in Partial Differential Equations, A.~Prastaro, Th.M.~Rassias eds.,
  World Scientific (1994), 155--172.
  
\bibitem{Magri:SMInHEq} F.~Magri, \emph{A simple model of the integrable
    {H}amiltonian equation}, J.\ Math.\ Phys. \textbf{19} (1978), 1156--1162.
  
\bibitem{Mar:FCM:2009} M.~Marvan, \emph{Sufficient set of integrability
    conditions of an orthonomic system}. Foundations of Computational
  Mathematics \textbf{9} (2009) 651--674.
  
\bibitem{Mikhailov:In} A.~Mikhailov (ed.), \emph{Integrability}, Lect.\ Notes
  Phys., vol.~767, Springer-Verlag, 2009.

\bibitem{NeyziNutkuSheftel:JPA:2005} F.~Neyzi, Y.~Nutku, and M.B.~Sheftel,
  \emph{Multi-Hamiltonian structure of Plebanski's second heavenly equation}
  J.\ Phys.\ A: Math.\ Gen.\ \textbf{38} (2005), 8473. \eprint{nlin/0505030}.

\bibitem{red} Obtaining \reduce: \url{http://reduce-algebra.sourceforge.net/}.
  
\bibitem{Vin:AMS} A.M.~Vinogradov , Cohomological Analysis of Partial
  Differential Equations and Secondary Calculus, vi+247~pp. Amer.\ Math.\
  Soc., Providence, RI (2001).

\bibitem{Vin:Found} A.M.~Vinogradov, \emph{On algebro-geometric foundations of
    {L}agrangian field theory}, Soviet Math.\ Dokl.\ \textbf{18} (1977),
  1200--1204.

\bibitem{Vin:Spectr} A.M.~Vinogradov, \emph{A spectral sequence associated
    with a nonlinear differential equation and algebro-geometric foundations
    of {L}agrangian field theory with constraints}, Soviet Math.\ Dokl.\
  \textbf{19} (1978), 144--148.

\bibitem{cdiff-man} R.~Vitolo, \emph{CDIFF: a REDUCE package for computations
    in geometry of differential equations}, the Geometry of Differential
  Equations website \url{http://gdeq.org} (2011).

\bibitem{WahlquistEstabrook:PSNEvEq} H.D.~Wahlquist and F.B.~Estabrook,
  \emph{Prolongation structures of nonlinear evolution equations}, J.\
  Mathematical Phys.\ \textbf{16} (1975), 1--7.

\bibitem{WahlquistEstabrook:PSNEvEqII} H.D.~Wahlquist and F.B.~Estabrook,
  \emph{Prolongation structures of nonlinear evolution equations. {II}}, J.\
  Mathematical Phys.\ \textbf{17} (1976), 1293--1297.
  
\bibitem{Zakharov:WIsIn} V.E.~Zakharov (ed.), \emph{What is integrability?},
  Springer-Verlag, 1991.
\end{thebibliography}
\end{document}